# Real-Time Tracking Antenna System for Moving Targets

Adham Saad[1], Aya Sherif Nassef[2], Mahmoud Mohamed Elshahed[3], Mohamed Ismail Ahmed[4,5]

[1,2,3]Communications and Information Engineering, Zewail City University of Science and Technology, Giza, Egypt
[4]RF Wireless and Mobile Systems Lab, Microstrip Dept., Electronics Research Institute, Cairo, Egypt
[5]Mechatronics Engineering Dept., Faculty of Engineering and Technology, Egyptian Chinese University, Cairo, Egypt, 11785
[1]adhamsaad@ieee.org, [2]ayanassef@ieee.org, [3]mahmoudelshahed@ieee.org, [4,5]miahmed@eri.sci.eg

*Abstract*—This paper presents the design and implementation of a compact, cost-effective phased array antenna system. It is capable of real-time beam-steering for dynamic target-tracking applications. The system employs a 4×4 rectangular microstrip patch array, utilizing advanced beamforming techniques and a Direction of Arrival (DoA) estimation algorithm. It achieves ±42° wide-angle scanning in both azimuth and elevation planes. The design emphasizes a balance between high angular coverage and consistent gain performance. This makes it suitable for wireless tracking, radar, and satellite communication terminals. Fabricated on Rogers 6010.2LM substrate, the system demonstrates reproducibility and scalability. All components are sourced locally to ensure practical deployment. The system is built using commercially available components, highlighting its affordability for research and prototyping purposes.

## A. System Overview and Motivation

Real-time beam tracking is vital for modern communication systems, particularly in radar, satellite terminals, and wireless sensing. This paper presents a compact, scalable phased array antenna system operating at 2.4 GHz in the ISM band. It supports dynamic 2D beam steering using Direction of Arrival (DoA) estimation and real-time beamforming.

Using electronic steering array (ESA) techniques, the array achieves ±42° scanning in both azimuth and elevation, ensuring wide angular coverage with stable gain. The system is built on Rogers 6010.2LM, chosen for its high dielectric constant and low loss tangent, enabling efficient high-frequency operation and array miniaturization. Full-wave simulations and MATLAB-based control validate the design's suitability for both lab prototypes and practical deployments.

*This work uniquely integrates wide-angle 2D beam steering and DoA-based signal processing in a compact, low-cost system built from off-the-shelf components—a combination not commonly achieved in prior literature at comparable size and cost*

## B. Literature Review

Several prior works have contributed significantly to real-time beam tracking for moving targets, offering valuable design insights.

Gogoi et al. [1] introduced a compact multimode radar operating in the 2.25–2.5 GHz band with 100 kHz resolution. Their design integrates RF, baseband, and power management modules on a small PCB and uses 5 dBi Vivaldi antennas. It supports both continuous-wave (CW) and frequency-modulated continuous-wave (FMCW) modes, making it suitable for short-range applications.

Kim et al. [2] developed a C-band planar active phased array with a beam steering range of ±45° in azimuth and 10° to 30° in elevation. By employing digital beamforming (DBF), their system achieves a high gain of 38.5 dBi with precise resolution and control.

Uddin et al. [3] demonstrated a phased array operating at 2.42 GHz with a beam steering range of 60° (1D) and a peak gain of 8.31 dBi. Their approach leverages spatial phase-shifting for improved resolution in radar applications.

Neumann et al. [4] presented a modular steerable phased array radar operating in the S-band (2.4–2.5 GHz). Their system provides a gain of 16 dBi at 2.45 GHz with a steering range of ±60° in both azimuth and elevation. The design's 200×200 mm compact footprint supports flexible deployment and precise calibration.

More recently, Zhang [5] proposed a dual-port, dual-beam pattern-reconfigurable antenna operating at 3.6 GHz. Their system supports independent 2D beam scanning with 400 MHz bandwidth and over 80% efficiency, offering a compact, high-performance solution for next-generation wireless systems.

## I. PROPOSED ANTENNA SYSTEM

### A. System Overview

The proposed system employs a 4×4 rectangular microstrip patch phased array, selected for its inherent beam steering capabilities and effectiveness in tracking moving targets. Rogers 6010.2LM substrate was selected for its high relative permittivity, which enables compact array dimensions, reduced inter-element spacing, and lower mutual coupling. This choice also aligns with practical constraints, as the required materials and fabrication facilities are available locally at the Electronics Research Institute (ERI), ensuring cost-effectiveness and feasibility within Egypt's infrastructure.

An inset feed technique was adopted for the feeding and matching network, preferred over quarter-wave transformers and coaxial feeds due to its impedance-matching efficiency and stable performance. For simulation purposes, a coaxial feed was used in CST to simplify the modeling and meet the development timeline of this conference paper.

Signal processing and visualization form an integral part of the system. A Direction of Arrival (DoA) algorithm estimates the target's angle, and the results are visualized through a custom MATLAB-based graphical user interface (GUI). Currently, the GUI demonstrates beam steering behavior and will be extended in future work to support real-time target tracking, offering insight into adaptive antenna systems.

TABLE I: Design Parameters of Unit Element Patch Antenna

| Parameter | Value | Unit |
|---|---|---|
| Resonant Frequency ($f$) | 2.4 | GHz |
| Wavelength ($\lambda$) | 125 | mm |
| Substrate Width ($W_s$) | 33.5 | mm |
| Substrate Length ($L_s$) | 26.6 | mm |
| Substrate Height ($H_s$) | 1.27 | mm |
| Patch Width ($W_p$) | 25.8 | mm |
| Patch Length ($L_p$) | 18.2 | mm |
| Metal Thickness ($T_s$) | 0.035 | mm |
| Length of Microstrip Feed Line ($L_f$) | 10.03 | mm |
| Width of Microstrip Feed Line ($W_f$) | 1.12 | mm |
| Cut Length ($L_{cut}$) | 8.1 | mm |
| Cut Width ($W_{cut}$) | 1.1 | mm |
| Element Spacing ($d$) | 50 (0.4$\lambda$) | mm |

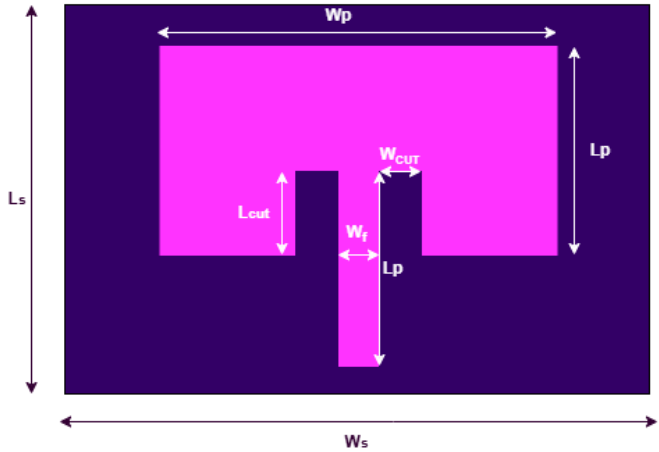

Fig. 1: Proposed Unit Cell Geometry

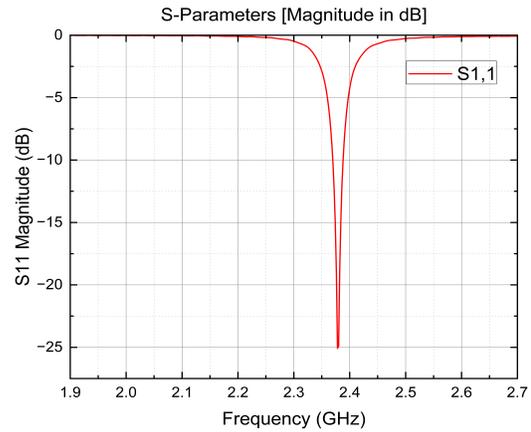

Fig. 3: $S_{11}$ Parameter Showing 2.4 GHz Resonance at -25 dB

As shown in Fig. 3, the simulated $S_{11}$ parameter of the unit cell exhibits a clear resonance around approximately 2.4 GHz. It achieves a return loss of -25 dB with a narrow 20 MHz bandwidth. This high frequency selectivity makes it suitable for precise target localization.

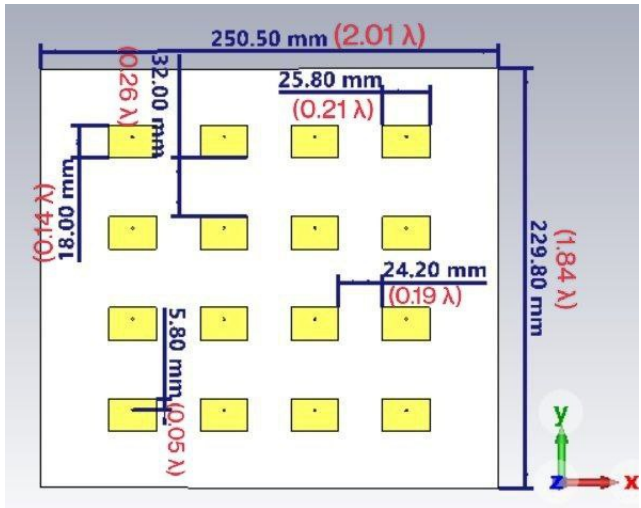

Fig. 2: 4x4 Antenna Array Configuration

The structure shown in Fig. 1 consists of a rectangular patch with a centrally etched T-shaped slot on a dielectric substrate, which controls the resonance and electromagnetic response. Fig. 2 presents the 16-element (4×4) phased array antenna with optimized spacing for beam-steering and grating lobe suppression. Key geometrical properties are summarized in Table I.

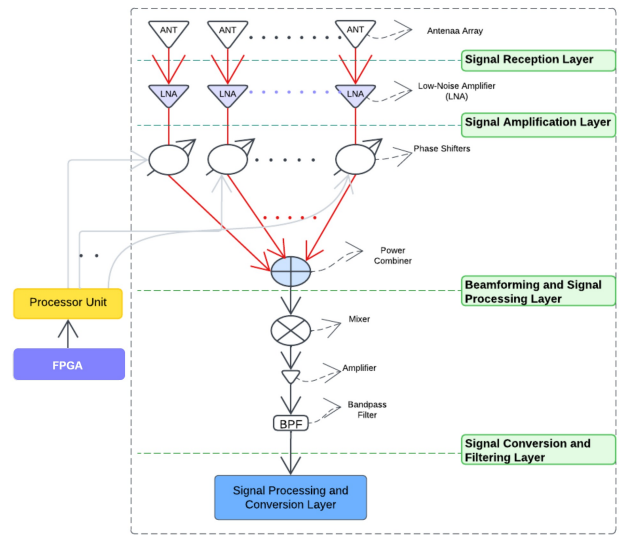

Fig. 4: System Block Diagram

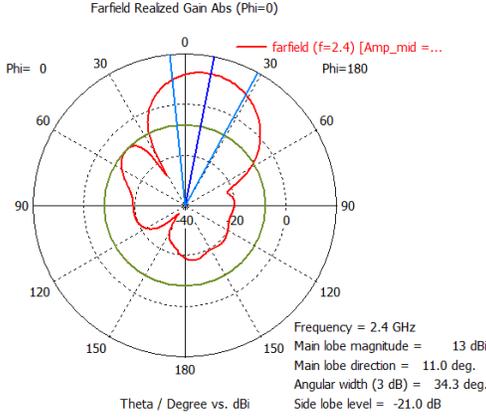
(a) **11° Steering:** 13 dBi gain, 34.3° beamwidth, -21.0 dB sidelobes.

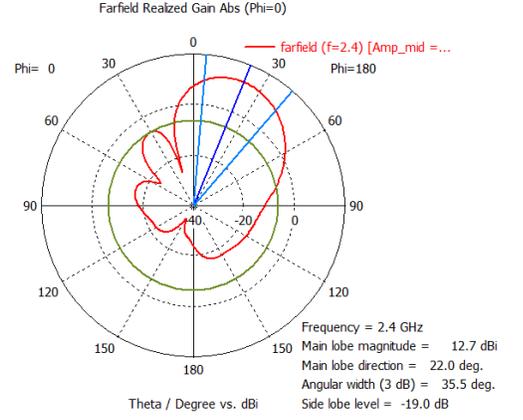
(b) **22° Steering:** 12.7 dBi gain, 35.5° beamwidth, -19.0 dB sidelobes.

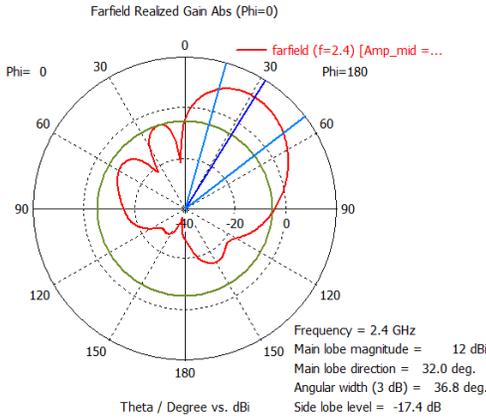
(c) **32° Steering:** 12 dBi gain, 36.8° beamwidth, -17.4 dB sidelobes.

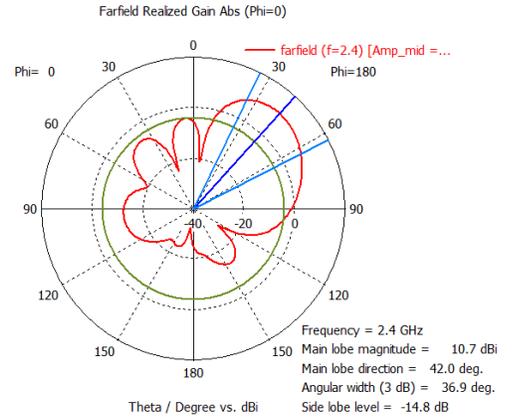
(d) **42° Steering:** 10.7 dBi gain, 36.9° beamwidth, -14.8 dB sidelobes.

Fig. 5: Far-field radiation patterns at 2.4 GHz showing beam steering at 11°, 22°, 32°, and 42°. Each plot indicates the main lobe gain, 3 dB beamwidth, and sidelobe suppression.

The system block diagram in Fig. 4 illustrates the architecture of the proposed antenna system for real-time target tracking. The signal flow progresses through reception, amplification, beamforming, and filtering stages. Beamforming is dynamically adjusted using an FPGA to maintain accurate target tracking. The final processed signal is delivered to the conversion layer for further analysis.

### B. Beam Tracking Capabilities

The proposed system achieves real-time beam tracking using a 4×4 phased array antenna with ±42° scanning angles in both azimuth and elevation.

**Theory of Operation:** Beam steering enables dynamic redirection of the antenna's main lobe toward desired signal sources. This capability enhances spatial selectivity, reduces interference, and supports accurate target tracking—making it essential for high-resolution sensing and wireless localization.

Table II compares related works. Unlike earlier studies with either narrow or 1D scanning, our antenna achieves wide 2D beam steering (±42°) with consistent 13.5 dBi gain. While some designs offer higher gain or smaller size, they sacrifice steering flexibility. Our larger footprint of 230×250 mm (1.84λ×2λ) supports high-gain scanning in both planes, addressing key gaps in prior solutions.

TABLE II: Literature Review of Phased Array Antenna Systems

| Ref. | Freq. Band (GHz) | Steering Angle | Gain (dBi) | Footprint (mm) |
|---|---|---|---|---|
| [1] | 2.25–2.5 | — | 5 | 80×80 (0.63λ×0.63λ) |
| [2] | 4–8 | ±45° | 38.5 | — |
| [3] | 2.42 | 60° (1D) | 8.31 | — |
| [4] | 2.4–2.5 | ±60° | 16 | 200×200 (1.63λ×1.63λ) |
| [5] | 3.6 | ±36° (2D) | 8.0 | 75×75 (0.9λ×0.9λ) |
| This Work | 2.4 | ±42° (2D) | 13.5 | 230×250 (1.84λ×2.0λ) |

## II. DIRECTION OF ARRIVAL (DoA) ESTIMATION

DoA estimation involves determining the angles of arrival (AoA) of signals reflected by a target. Using the Multiple Signal Classification (MUSIC) algorithm, the azimuth and elevation angles of the target can be estimated with high resolution. This section outlines the process and presents the simulation results.

### A. MUSIC Algorithm Overview

To determine the Angle of Arrival (AoA) of a target, a 2.4 GHz signal is first transmitted toward the target, and the reflected signal is captured using a rectangular array, with attention to phase differences across the array elements. The covariance matrix of the received signals is then computed, followed by eigenvalue decomposition to isolate the signal and noise subspaces. Using these, the Multiple Signal Classification (MUSIC) spectrum is generated by scanning azimuth ($\theta$) and elevation ($\phi$) angles, where prominent peaks in the spectrum correspond to the target's AoA.

### B. Simulation Results

MATLAB simulations provided the following results:
- True Angles: Azimuth $\theta$: $-30°$, Elevation $\phi$: $22.9389°$.
- Estimated Angles: Azimuth $\theta$: $20°$, Elevation $\phi$: $-34°$. The results demonstrate the MUSIC algorithm's potential but reveal discrepancies between true and estimated angles, indicating areas for improvement in accuracy.

### C. Visualization of Results

Figures 7 and 6 show the estimated angles compared to the true target angles:

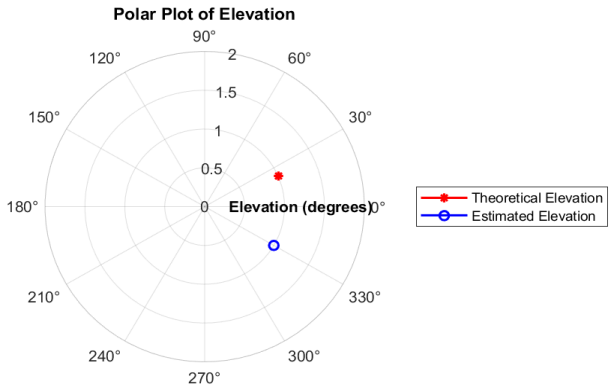

Fig. 6: Elevation angle estimation: True vs. Estimated.

The polar plot, as seen in Fig. 6, compares the theoretical and estimated elevation angles of the target, demonstrating good estimation performance with a slight shift in the estimated angle.

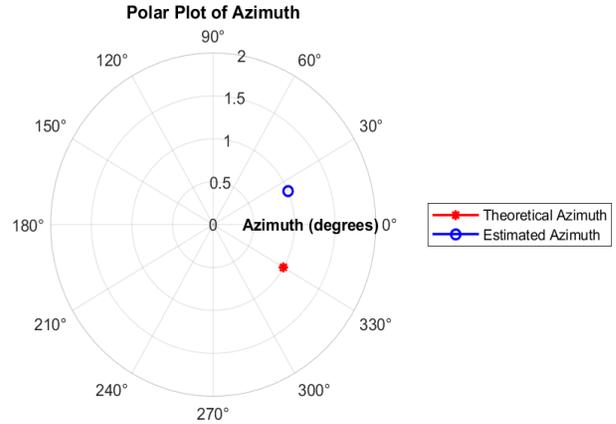

Fig. 7: Azimuth angle estimation: True vs. Estimated.

The polar plot, as seen in Fig. 7, compares the theoretical and estimated azimuth angles of the target. The plot demonstrates good estimation performance with a slight shift in the estimated angle.

### D. Error Analysis and Improvement Suggestions

The observed mismatch between the true and estimated angles in the DoA results stems from several key limitations in the current simulation setup. Estimation is performed on a single snapshot, which reduces the robustness of the covariance matrix, especially under the influence of noise—even at 20 dB SNR. The use of a fixed 1° angular resolution may also introduce quantization errors, particularly when the true angles fall between scan points. Finally, the absence of array calibration can result in phase and gain mismatches across elements, further degrading accuracy. To mitigate these issues, improvements such as array calibration, temporal averaging, multipath mitigation, and using finer angular steps in the scan can significantly enhance estimation accuracy.

## III. COST ANALYSIS

The proposed phased array antenna system was developed with cost-effectiveness as a primary design consideration. The total system cost is approximately $1000, covering all essential components including the 4×4 microstrip patch array, RF transceiver, amplifiers, filters, beamforming module, FPGA controller, power supply units, and supporting hardware. High-cost components—such as the RF amplifiers and beamforming modules—were necessary to meet the performance targets, but overall expenses were minimized by careful integration and the use of widely available commercial off-the-shelf (COTS) parts.

## IV. USER INTERFACE

The proposed system includes a MATLAB-based user-friendly interface for visualizing beam-tracking in real time. This simple UI enables ease of visualization for non-specialists as seen in Fig. 8.

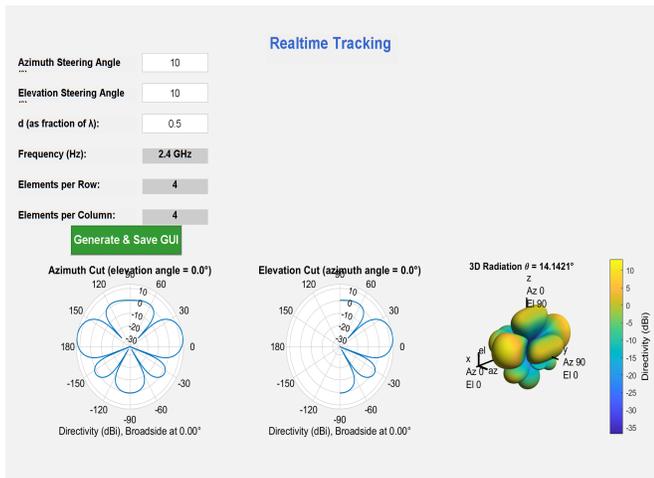

Fig. 8: Graphical User Interface for real-time beam tracking visualization.

## V. Conclusion

This paper presented a compact and cost-effective 4×4 phased array antenna system capable of real-time 2D beam steering for dynamic target tracking. The system achieves ±42° scanning in both azimuth and elevation planes while maintaining stable gain, validated through full-wave simulations and MATLAB-based control. A key contribution of this work is the integration of wide-angle 2D beam steering with Direction of Arrival (DoA)-based signal processing in a low-cost, compact implementation. This combination of angular coverage, gain stability, and implementation simplicity is rarely demonstrated in prior literature at comparable scale and resource constraints. The array is fabricated on Rogers 6010.2LM substrate for high efficiency and reduced size, and all components were locally sourced to ensure feasibility, reproducibility, and cost-effectiveness. A hardware prototype is planned as future work to evaluate real-time performance. Overall, the proposed system offers a practical and scalable solution for beam-tracking applications, with strong potential for deployment in next-generation radar and wireless communication platforms.

## References

[1] P. K. Gogoi, M. K. Mandal, A. Kumar, and T. Chakravarty, "A Compact Multi-mode Integrated Doppler radar at 2.4 GHz for Multipurpose Applications," *2022 IEEE Wireless Antenna and Microwave Symposium (WAMS)*, Rourkela, India, 2022, pp. 1–4, doi: 10.1109/WAMS54719.2022.9848033.

[2] K.-H. Kim *et al.*, "Development of planar active phased array antenna for detecting and tracking radar," *2018 IEEE Radar Conference (RadarConf18)*, Oklahoma City, OK, USA, 2018, pp. 0100–0103, doi: 10.1109/RADAR.2018.8378538.

[3] N. ud Din, A. Afzaal, J. Tahir, and N. Hafeez, "Electronic beam scanning for radar applications," *8th International Conference on High-capacity Optical Networks and Emerging Technologies*, Riyadh, Saudi Arabia, 2011, pp. 355–358, doi: 10.1109/HONET.2011.6149769.

[4] N. Neumann, C. Hammerschmidt, M. Laabs, and D. Plettemeier, "Modular steerable active phased array antenna at 2.4 GHz," *2016 German Microwave Conference (GeMiC)*, Bochum, Germany, 2016, pp. 333–336, doi: 10.1109/GEMIC.2016.7461624.

[5] Y. Zhang, S. Tang, J. Rao, C. -Y. Chiu, X. Chen and R. Murch, "A Dual-Port Dual-Beam Pattern-Reconfigurable Antenna With Independent 2-D Beam-Scanning," in IEEE Transactions on Antennas and Propagation, vol. 72, no. 10, pp. 7628-7643, Oct. 2024, doi: 10.1109/TAP.2024.3452425.

[6] Y.-J. Yang, Y.-L. Chen, B.-A. Liu, J.-H. Chen, and Y.-J. E. Chen, "Development of 2.4 GHz phased array transmitter for wireless power transfer," 2024 IEEE Wireless Power Transfer Conference (WPTCE), Kyoto, Japan, 2024, pp. 1–4, doi: 10.1109/WPTCE59894.2024.10557329.

[7] Y. S. Rusov, D. R. Russo, and P. P. Kurenkov, "Development of a beam steering system for a phased antenna array with variable duration of control pulses," 2024 6th International Youth Conference on Radio Electronics, Electrical and Power Engineering (REEPE), Moscow, Russian Federation, 2024, pp. 1–4, doi: 10.1109/REEPE60449.2024.10479954.

[8] R. C. Hansen, Phased Array Antennas, 2nd ed. Hoboken, NJ, USA: John Wiley Sons, Inc., 2009, pp. 7–46, 129–216.

[9] C. A. Balanis, Antenna Theory: Analysis and Design, 3rd ed. Hoboken, NJ, USA: John Wiley & Sons, Inc., 2005, pp. 283–371.

[10] S. J. Orfanidis, Electromagnetic Waves and Antennas, Piscataway, NJ, USA: Sophocles J. Orfanidis, 1999–2016, pp. 1088–1116.

[11] R. Garg, Microstrip Antenna Design Handbook, Boston, MA, USA: Artech House, 2001, pp. 253, 317–719, 759.

[12] D. Buck, "Design and characterization of phased arrays for UAS detection and tracking," Brigham Young University ScholarsArchive, Aug. 2, 2022. [Online]. Available: https://scholarsarchive.byu.edu/cgi/viewcontent.cgi?article=10724&context=etd. [Accessed: Dec. 28, 2024].

[13] Ö. Ipek, "Target tracking with phased array radar by using adaptive update rate," OpenMETU. [Online]. Available: https://open.metu.edu.tr/handle/11511/19367. [Accessed: Dec. 28, 2024].

[14] MathWorks, "Scan radar using a uniform rectangular array," *MATLAB Phased Array System Toolbox User's Guide*, [Online]. Available: https://www.mathworks.com/help/phased/ug/scan-radar-using-a-uniform-rectangular-array.html. [Accessed: Dec. 28, 2024].